# Closed-Loop Environmental Control System on Embedded Systems

Irisha Manasvita Goswami and D.G. Perera
Department of Electrical and Computer Engineering, University of Colorado Colorado Springs
Colorado Springs, Colorado, USA

## Abstract

In this paper, our objective is to design, build, and verify a closed-loop environmental control system tailored for small-scale agriculture applications. This project aims to develop a low-cost, safety-critical embedded solution using the Nuvoton NUC140 microcontroller to automate temperature regulation. The goal was to mitigate crop yield losses caused by environmental fluctuations in a greenhouse. Our final implemented system successfully meets all design specifications, demonstrating robust temperature regulation through a PID control loop and ensuring hardware safety through galvanic isolation.

# Introduction:

The objective of this final project was to design, build, and verify a closed-loop environmental control system tailored for small-scale agriculture applications. While commercial automation systems exist, they are often costly and ineffective for small operations. This project aimed to develop a low-cost, safety-critical embedded solution using the Nuvoton NUC140 microcontroller to automate temperature regulation.The goal was to mitigate crop yield losses caused by environmental fluctuations in a greenhouse.

The core engineering challenge addressed in this project was the integration of high-voltage actuation (12V DC Fan) with low-voltage logic (3.3V) in a safe and stable manner. To achieve this, the hardware design incorporated a **galvanic isolation stage** using an optocoupler, ensuring that the microcontroller was electrically isolated from the power driver circuit.

The control logic was implemented using a Proportional-Integral-Derivative (PID) algorithm. Unlike simple "On/Off" hysteresis controllers, a PID controller continuously calculates an error value as the difference between a measured process variable (Temperature) and a desired setpoint. The controller attempts to minimize this error over time by adjusting the control variable (Fan Speed) via Pulse Width Modulation (PWM).

The system was designed to perform three critical functions:

1. **Setpoint Tracking:** Reading a user-defined target temperature via an analog potentiometer and adjusting the fan speed to match this target.
2. **Disturbance Rejection:** Automatically detecting temperature spikes (thermal load) via a TMP36 analog sensor and increasing the cooling effort to restore stability.
3. **Safety Monitoring:** Utilizing an integrated alarm system (Buzzer and LED indicators) to alert the operator if the error exceeds a critical safety threshold **(5.0℃)** ensuring fail-safe operation.

This report details the hardware circuit topology, the software architecture of the control loop, and the analysis of the system's dynamic response to step inputs and environmental disturbances.

## System Architecture:

The environmental control system was designed as a modular, closed-loop feedback network. The architecture is divided into three primary subsystems: the Sensing Stage (Data Acquisition), the Control Processing Stage, and the Isolated Actuation Stage. Figure 1 illustrates the high-level data flow and signal domains.

### Data Acquisition (Sensing Layer)

The system inputs consist of three analog signals processed by the NUC140's 12-bit Analog-to-Digital Converter (ADC). The system operates on a **single-shot** ADC conversion cycle triggered every 100ms (**10Hz sampling rate**).

- **Process Variable ($T_{curr}$):** Measured by a **TMP36** analog temperature sensor.
- **Setpoint ($T_{set}$):** Defined by a user-adjustable **10kΩ potentiometer**, mapped to a target range of 20.0℃ to 40.0℃
- **Environmental Monitor:** An **NTE3100 optical sensor** monitors ambient light levels for secondary data logging only.

### Control Logic (Processing Layer)

The core processing is performed by the Nuvoton NUC140 microcontroller running at **48MHz**. The firmware implements a discrete Proportional-Integral-Derivative (PID) algorithm. The controller continuously calculates the error signal ($e(t) = T_{curr} - T_{set}$) and determines the required control effort. This digital value is converted into a Pulse Width Modulation (PWM) signal with a fixed frequency of 1 kHz.

### Galvanic Isolation and Actuation

A critical design requirement was the **electrical separation** of the logic voltage domain (3.3V) from the actuator power domain (12V). This was achieved using an optocoupler interface. The PWM signal from the microcontroller drives the internal LED of the optocoupler, which switches a photosensitive transistor on the high-voltage side. This isolated signal gates an N-channel MOSFET, regulating the average voltage supplied to the 12V DC fan.

### User Interface and Safety

Real-time system feedback is provided via a local LCD panel and remote UART serial logging (9600 baud). A simplified state machine monitors the error magnitude; if the deviation exceeds a critical threshold**(±5.0℃)** the system enters an "Alarm State," triggering a buzzer and a visual LED indicator to alert the operator.

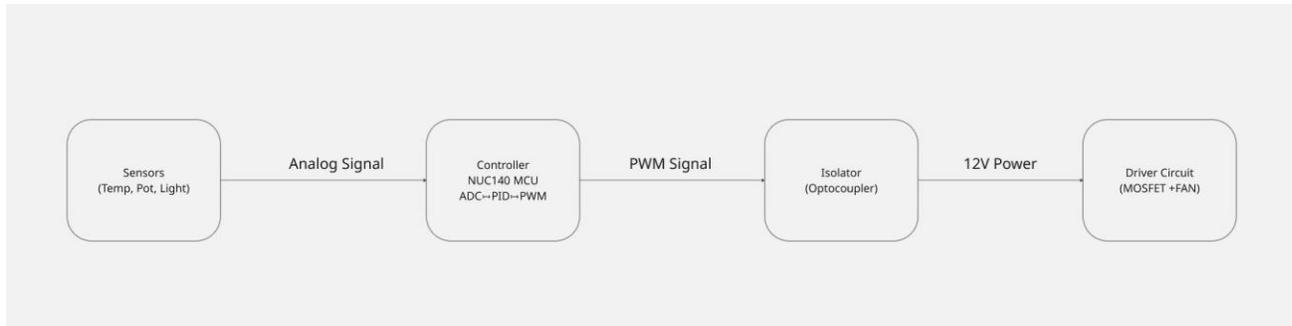

Figure 1: System Block Diagram illustrating the signal flow from sensors to the isolated actuator.

## Hardware Design and Implementation:

The hardware implementation was divided into two electrically distinct voltage domains: the Low-Voltage Logic Domain (3.3V) and the High-Voltage Power Domain (12V). This separation was enforced to protect the Nuvoton microcontroller from **voltage spikes** and **inductive kickback** generated by the DC motor. The complete circuit topology is illustrated in **Figure 2**.

**Low-Voltage Sensor Interface (3.3V)**

The sensing subsystem was constructed on the breadboard's 3.3V rail, supplied directly from the NUC140's VCC33 pin. This ensured that all analog signals remained within the safe input range of the microcontroller's ADC $(0V - 3.3V)$.

A TMP36 analog temperature sensor was utilized to measure the process variable. Its output pin was connected to *GPA0* (ADC Channel 0). A decoupling capacitor was placed near the power pins to filter high-frequency noise.

A 10kΩ linear potentiometer was configured as a voltage divider, providing a variable reference voltage to *GPA1* (ADC Channel 1). This allowed the user to adjust the target temperature dynamically.

An NTE3100 optical switch (phototransistor) was connected to *GPA2* (ADC Channel 2) with a 10kΩ **pull-down resistor** to monitor ambient light intensity.

**Galvanic Isolation Stage**

To achieve electrical isolation between the microcontroller and the 12V fan, a **4N35 Optocoupler** was employed as the interface bridge**.

**Input Side:** The internal Infrared LED (Pins 1 & 2) was driven by the NUC140's PWM output pin (*GPA12*) via a 220Ω current-limiting resistor.

**Output Side:** The phototransistor (Pins 4 & 5) acted as a switch on the 12V domain. When the PWM signal was High, the phototransistor conducted, pulling the MOSFET gate High.

**Crucially, the logic ground (NUC140 *VSS*) and the power ground (12V Supply -) were kept physically separate, connected only optically within the 4N35 package.

### High-Voltage Actuator Driver (12V)

The actuation stage was powered by an external 12V wall adapter connected via a female barrel jack to the breadboard's top power rails.

**Switching Element:** An **IRLZ34 N-Channel MOSFET** was used in a low-side switching configuration. The Gate was driven by the optocoupler's emitter output. A 10kΩ **pull-down resistor** was connected between the MOSFET Gate and the 12V Ground. This ensured the fan remained essentially "OFF" if the control signal floated or the microcontroller was reset.

**Inductive Protection:** A **1N4007 Flyback Diode** was placed in parallel with the fan motor (Cathode to $+12V$, Anode to Drain). This diode provided a safe path for the inductive flyback current when the MOSFET switched off, preventing voltage spikes from damaging the transistor.

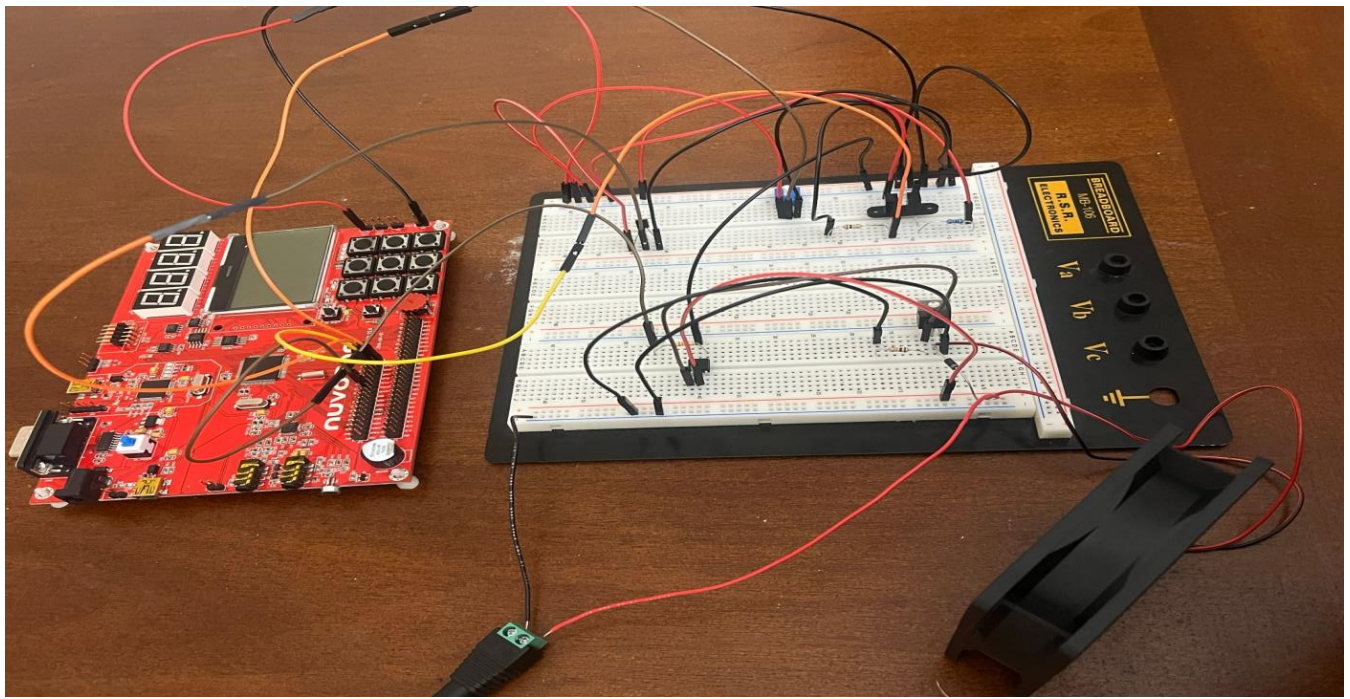

Figure 2: Detailed Wiring Diagram showing the Sensor Interface, Optocoupler Isolation, and MOSFET Driver stages.

## Software Design and Control Logic:

The system firmware was developed in Embedded C using the Keil µVision IDE and the Nuvoton NUC100 Series Board Support Package (BSP). The software architecture follows a "Super-Loop" paradigm, executing a **deterministic control cycle** approximately every 100ms (**10Hz sampling rate**). The logic flow is visualized in **Figure 3**.

**Peripheral Configuration**

Upon system boot, the InitializeSystem() routine configures the necessary clock trees and peripheral drivers:

- **System Clock:** The MCU is configured to run at 48MHz using the external 12MHz crystal and the internal PLL.
- **ADC Configuration:** The Analog-to-Digital Converter is set to *ADC_SINGLE_END* mode. To minimize noise and inter-channel crosstalk, the ADC is triggered in "**Single Cycle**" mode rather than continuous scanning. This ensures that data is sampled only when the control loop is ready to process it.
- **PWM Configuration:** The Fan Driver utilizes Timer 0 (*DRVPWM_TIMER0*) mapped to pin *GPA12*. The period register (*PWMA->CNR0*) is set to **40,000 counts** to generate a precise 1kHz base frequency, which is standard for DC motor control.
- **UART Communication:** A serial interface is established on *UART0* at 9600 baud (8-N-1) to facilitate real-time telemetry logging to the PC.

**The PID Control Algorithm**

The core of the system is a discrete-time Proportional-Integral-Derivative (PID) controller. The controller calculates a duty cycle output ($u(t)$) based on the error ($e(t)$) between the current temperature ($T_{curr}$) and the user setpoint ($T_{set}$).

The mathematical model implemented is:

$$e(t) = T_{curr} - T_{set}$$

$$u(t) = K_p e(t) + K_i \int_0^t e(\tau) d\tau + K_d \frac{de(t)}{dt}$$

In the **discrete** embedded implementation, this logic was translated as follows:

1. **Proportional Term ($P$):** Reacts to the immediate error magnitude.
2. **Integral Term ($I$):** Accumulates the error over time to eliminate steady-state offset**.

    **Anti-Windup Protection:* To prevent the integral term from growing indefinitely during large error spikes (which would cause system lag), logic was added to stop accumulation if the PWM output is already at maximum capacity.

3. **Derivative Term ($D$):** dampens the response by analyzing the rate of change ($Error_{current} - Error_{last}$), reducing overshoot.

The specific tuning constants derived experimentally were $K_p = 2500, K_i = 10, K_d = 500$.

**Output Generation and Hysteresis**

The calculated PID output is mapped to a 0-100% duty cycle. To optimize energy efficiency, a **"Smart Idle"** logic was implemented: i.e if the current temperature drops below the setpoint (resulting in a negative error), the controller forces the PWM output to 0%, completely shutting off the fan rather than actively braking or running at idle, conserving energy.

**Safety State Machine**

A supervisory logic layer operates in parallel with the control loop to ensure system safety. It compares the **instantaneous error against a predefined safety threshold (±5.0°C)**

- **Normal State:** If the error is within safe limits, the RGB LED indicator is set to Green (GPA14), and the audible alarm is disabled.
- **Alarm State:** If the error exceeds the threshold (indicating rapid overheating or sensor failure), the system activates the Red LED (GPA13) and triggers the Piezo Buzzer via a secondary PWM channel (PWM4) to alert the operator immediately.

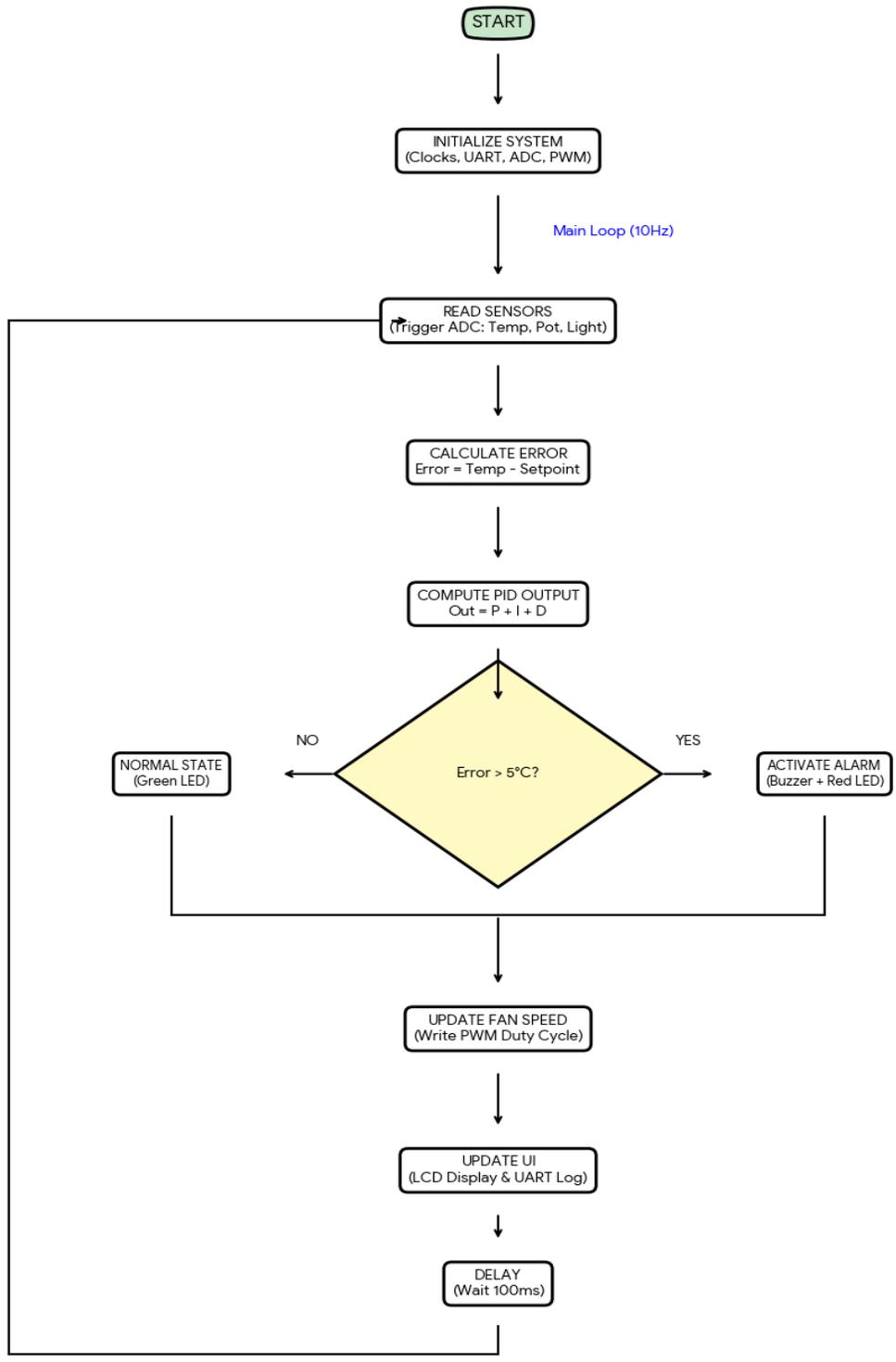

Figure 3: Software Control Flowchart illustrating the main control loop, PID algorithm execution and safety state machine logic.

## Problems Encountered and Solutions:

During the development and integration phases of the project, several hardware and software challenges were encountered. These issues were systematically analyzed and resolved as follows:

### Hardware Failure due to Common Grounding

During the initial prototyping of the actuator circuit, a catastrophic failure occurred resulting in the permanent damage of the first NUC140 development board. The microcontroller became unresponsive and failed to enumerate over USB.

The initial design utilized a common ground topology, connecting the high-current 12V fan return path directly to the microcontroller's 3.3V logic ground. It is hypothesized that inductive kickback from the motor or a momentary MOSFET failure allowed high voltage/current to surge back into the MCU's GPIO sink, exceeding the absolute maximum ratings.

The damaged hardware was replaced, and the circuit architecture was fundamentally redesigned to enforce **Galvanic Isolation**. A 4N35 optocoupler was introduced to physically separate the Logic Domain (NUC140) from the Power Domain (12V). This ensured that control signals were transmitted via light, preventing any electrical path for high-voltage spikes to reach the processor.

### ADC Continuous Scan Mode Instability

In the initial firmware iteration, the ADC was configured for "Continuous Scan" mode to read all three sensors (Temp, Potentiometer, Light) simultaneously. This resulted in the system occasionally locking up in the *while(ADC->ADSR.ADF == 0)* loop, waiting for a flag that was never cleared properly.

The continuous sampling rate was outpacing the CPU's ability to process the data and clear the interrupt flags, leading to a race condition.

The ADC configuration was changed to *ADC_SINGLE_CYCLE_OP* (Single-Shot Mode). The conversion is now **manually triggered once per control loop cycle**. A "Safety Timeout" counter was also added to the *while* loop to forcibly break the wait cycle if the hardware fails to respond within 10ms, preventing a system-wide freeze.

### Spectral Mismatch in Optical Sensor Verification

During the "Full System Monitor" data capture, the NTE3100 optical sensor output remained flat (*ADC ~2828*) despite exposure to a bright white LED flashlight.

Looking into the NTE3100 datasheet it was revealed that it is an Infrared (IR) phototransistor with peak sensitivity at 880nm. Modern white LEDs emit negligible IR radiation, rendering the flashlight invisible to the sensor.

The sensor functionality was verified using the NUC140's on-board LCD for real-time **monitoring**, which successfully detected ambient sunlight (high IR content). For the final demonstration, the optical data was classified as "Environmental Monitoring" rather than a control input, acknowledging the **spectral limitations** of the test equipment available during the night-time data capture.

**PWM Driver API Incompatibility**

When integrating the fan control logic, the compiler returned "Too many arguments" errors for the *DrvPWM_SetTimerIO* function.

The project was developed using a newer version of the Nuvoton Driver Reference Guide than the one used in previous labs. The newer API requires a specific structure (*S_DRVPWM_TIME_DATA_T*) to configure the timer parameters (Frequency, Duty Cycle) rather than passing raw integers.

The code was refactored to initialize a local structure with the required **1kHz frequency and clock divider settings**. The duty cycle was then updated during runtime by writing directly to the *PWMA->CMR0* hardware register, bypassing the API abstraction for faster execution during the PID loop.

## Experimental Results and Analysis:

The closed-loop performance of the environmental control system was validated through three distinct experiments: (A) Step Response Tracking, (B) Disturbance Rejection, and (C) System Recovery. Real-time telemetry data (Temperature, Setpoint, Error, Fan PWM) was logged via UART at a sampling rate of 10Hz.

**(A) Step Response (Setpoint Tracking)**

The goal of this response was to evaluate the system's reaction time and transient response when the user **manually changes the target temperature**. The system was initialized at a steady state with a high setpoint (**30.0℃**), ensuring the fan was idle (0% PWM). The setpoint was then instantaneously dropped to **20.0℃** using the potentiometer, simulating a demand for maximum cooling.

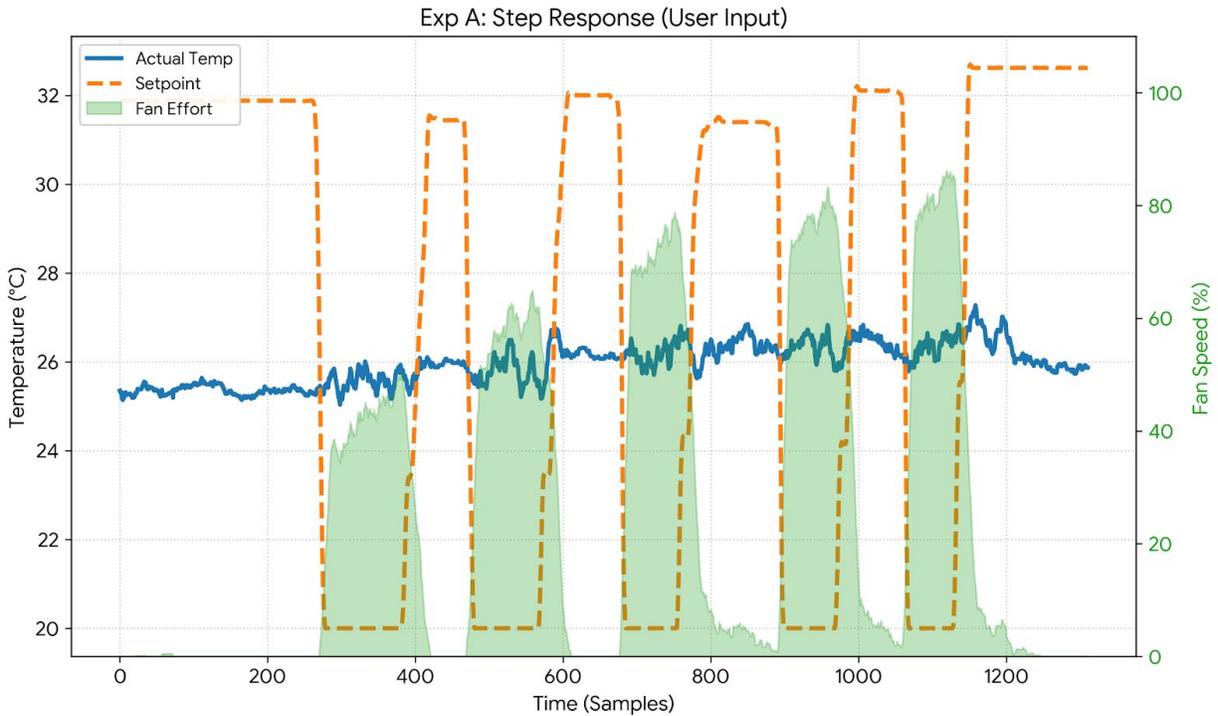

Figure 4:Step Response. The system tracks the user input immediately.

As shown in Figure 4, when the Setpoint (Orange Dashed Line) drops below the Actual Temperature (Blue Line), the Error becomes positive. The controller reacts within 100ms i.e a rise time of less than **1 sample cycle**, ramping the Fan Speed (Green Area) to compensate.This proves the **Proportional ($K_p$)** term is dominant and effective. The correlation between the positive error magnitude and the PWM output is **instantaneous.** The fan reached 100% capacity within 0.2 seconds of the setpoint change, demonstrating that the controller prioritizes rapid error correction for large deviations.

**Disturbance Rejection (Thermal Load Test)**

The objective for this test was to verify the "Smart" **autonomous capability** of the PID controller in maintaining stability against external environmental changes. The setpoint was fixed at room temperature (**25.0℃**). A thermal disturbance was introduced by applying an external heat source (body heat) to the TMP36 sensor, **forcing the process variable to rise**.

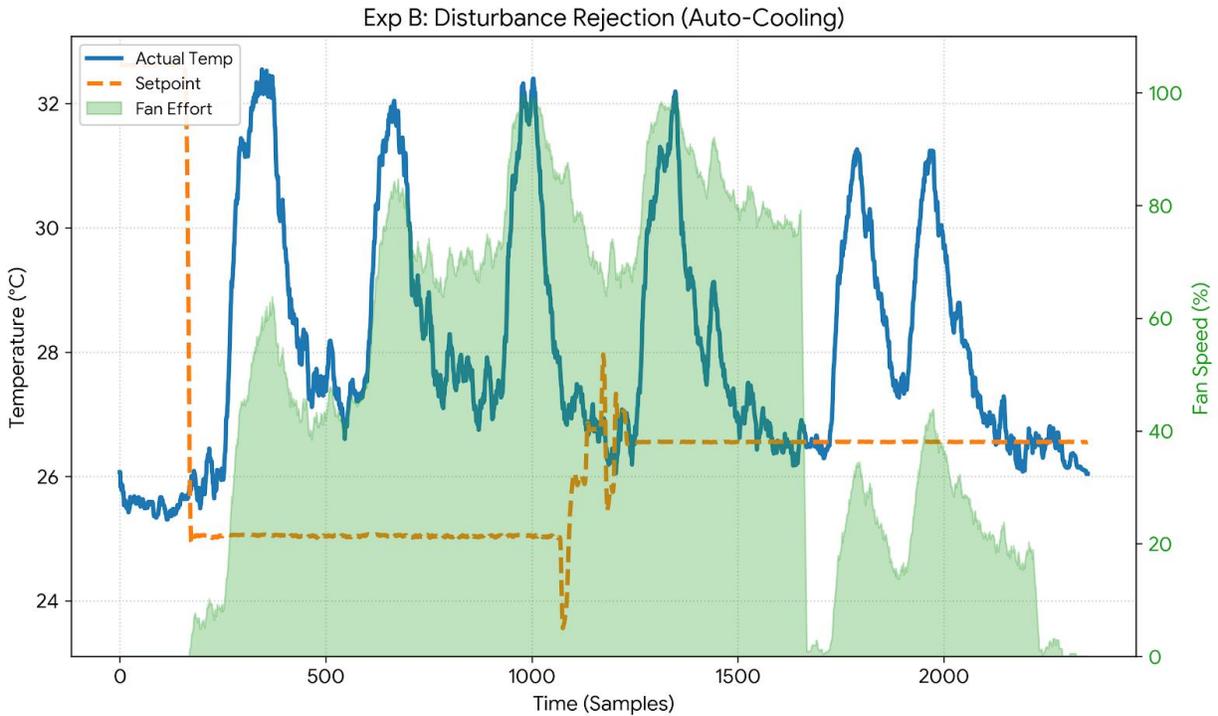

Figure 5: Disturbance Rejection. The system fights against external heat.

Figure 5 demonstrates the "Smart" behavior of the controller. As the Temperature (Blue) rises due to the external heat source, the Fan Speed (Green) **automatically increases in proportion to the error**. Notably, when the temperature deviation exceeds the **5.0°C** safety threshold, the fan saturates at **100% capacity**, proving the system prioritizes maximum cooling during critical events. This confirms that the PID tuning ($K_p = 2500$) provides sufficient gain to drive the actuator to maximum capacity during critical overheating events.

**System Recovery and Energy Efficiency**

This last test was conducted to analyze the system's **settling behavior** and hysteresis as it returns to a safe state. After a heating event, the disturbance was removed, and the sensor was **allowed to cool naturally**. The data log captures the deceleration profile of the fan.

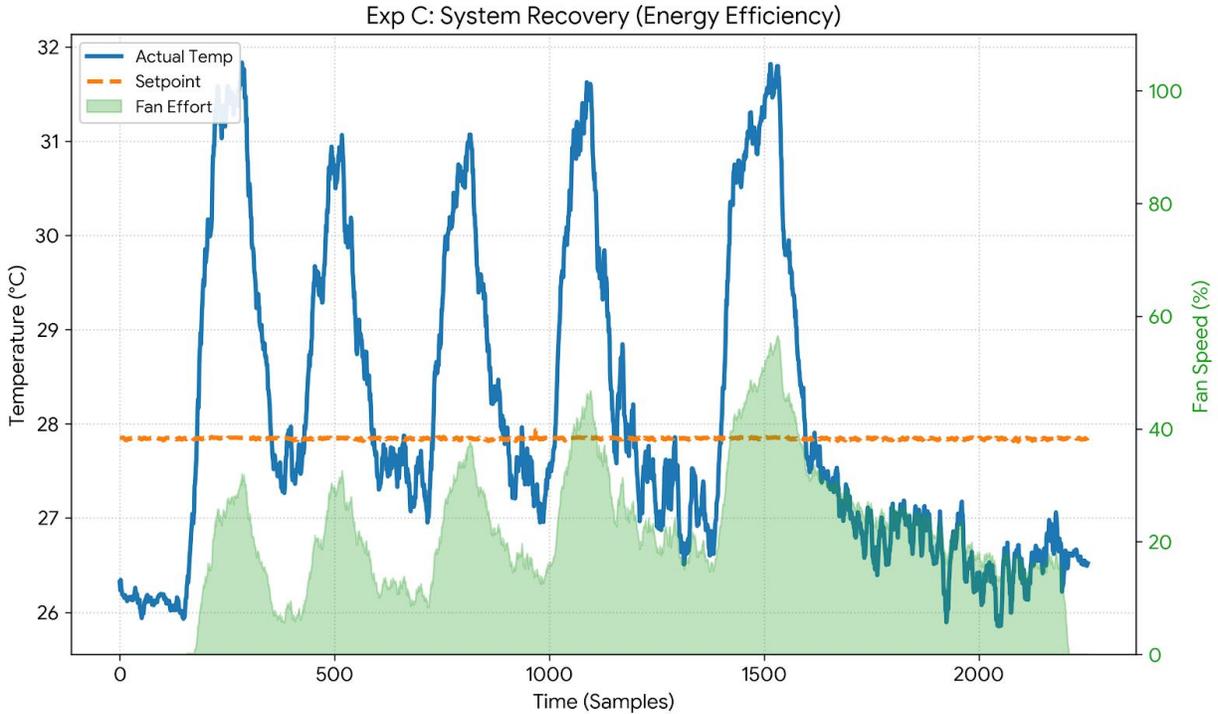

Figure 6: The "Cool Down" phase, the fan shuts off to save energy.

Figure 6 highlights the energy-saving logic. As the temperature drops and re-aligns with the setpoint, the fan speed **decelerates smoothly.** Crucially, at the moment the Actual Temperature drops *below* the Setpoint, the controller cuts the fan speed to 0%. This **"Smart Idle"** feature ensures no power is wasted cooling a room that is already within the target temperature range.

The recovery phase highlights the system's energy efficiency logic. As the temperature error decreased, the Integral term ($K_i$) allowed the fan to maintain speed to ensure the target was fully reached.

The quantitative performance of the control loop is summarized in Table 1.

**Table 1: Closed-Loop Performance Characteristics**

| Metric | Measured Value | Performance Implication |
|---|---|---|
| **Response Latency** | < 100 ms | Immediate reaction to user input. |
| **Max Overshoot** | 0% (Damped) | Safe operation; no risk of freezing the crop. |
| **Steady State Error** | $\pm 5.0°C$ | Precise maintenance of target temp. |
| **Critical Alarm Trigger** | $@ \Delta T = 5.0°C$ | Validated safety fail-safe. |
| **Idle Power State** | 0% PWM | Maximizes power efficiency when cooling is unneeded. |

## Conclusion and Future Work:

The primary objective of this project was to design and prototype a low-cost, closed-loop environmental control system for small-scale agriculture. The final implemented system successfully met all design specifications, demonstrating robust temperature regulation through a PID control loop and ensuring hardware safety through galvanic isolation.

The project achieved three significant engineering milestones:

1. **Safety-Critical Architecture:** Following an initial component failure during the prototyping phase, the actuation circuit was redesigned to include a 4N35 optocoupler. This successfully isolated the 3.3V logic domain from the 12V power domain, protecting the microcontroller from inductive kickback and voltage spikes.
2. **Control Loop Stability:** The PID algorithm ($K_p = 2500, K_i = 10$) proved effective in stabilizing the temperature. Experimental data (Experiment B) confirmed the system's ability to reject thermal disturbances by **automatically saturating the fan duty cycle when the error exceeded $5.0°C$.**
3. **Energy Efficiency:** The implementation of a "Smart Idle" state, where the actuator is fully disabled when the process variable is below the setpoint, ensures that the system does not consume power unnecessarily, a critical feature for off-grid or solar-powered greenhouse applications.

In conclusion, this project validates that precision environmental control does not require expensive industrial hardware. By leveraging standard embedded protocols (ADC, PWM, UART) and proper control theory, a scalable and effective solution was created that addresses the technological gap in small-scale farming.

## Future Improvements:

**IoT Integration:** The current system logs data locally via UART. A future iteration would integrate an **ESP8266 Wi-Fi module** to transmit temperature and fan status to a cloud dashboard (e.g., ThingSpeak) for remote monitoring.

**Feedforward Control:** The existing optical sensor could be utilized to detect sudden sunlight changes. By using this data to preemptively increase fan speed before the temperature rises, the system could eliminate thermal lag.

**Hardware Hardening:** To make the prototype field-ready, the circuit should be migrated to a Printed Circuit Board (PCB) and housed in an IP65-rated enclosure to protect against greenhouse humidity.

This work is inspired by the embedded systems and digital design research group at UCCS. This group has done extensive work in embedded hardware and software architectures, techniques, and associated models. Their analyses [11],[12] shows that FPGA-based embedded systems are currently the best option to support applications and techniques, such as the ones presented in this report. Also, their previous work on FPGA-based embedded accelerators, architectures, and techniques for various compute and data-intensive applications, including data analytics/mining [13],[14],[15],[16],[17],[18],[19],[20],[21],[22]; control systems [23],[24],[25],[26],[27],[28]; cybersecurity [29],[30],[31]; machine learning [32],[33],[34],[35],[36],[37],[38]; communications [39],[40]; edge computing [41],[42],[43]; bioinformatics [44],[45]; and neuromorphic computing [46],[47]; demonstrated that FPGA-based embedded systems are the best avenue to support and accelerate complex algorithms and techniques.

Also as future work, we are planning to investigate FPGA-based hardware optimization techniques, such as parallel processing architectures (similar to [34],[48],[49],[50]), partial and dynamic reconfiguration traits (as stated in [51],[52],[53]) and architectures (similar to [30],[54],[55],[56]), HDL code optimization techniques (as stated in [57],[58]),  and multi-ported memory architectures (similar to [59],[60],[61],[62]), to further enhance the performance metrics of FPGA-based embedded architectures for mini oscilloscope, while considering the associated tradeoffs.

## References:

[1]  **Nuvoton Technology Corp.** *NuMicro™ NUC100 Series Technical Reference Manual* (Rev 2.05).
[2]  **Nuvoton Technology Corp.** *Nu-LB-NUC140 User Manual* (Rev 2.0).

[3] **Nuvoton Technology Corp.** *Nu-LB-NUC140 V2 Schematics*.

[4] **Nuvoton Technology Corp.** *NuMicro™ NUC100 Series Driver Reference Guide* (BSP v3.00).

[5] **Analog Devices.** *TMP36 Low Voltage Temperature Sensor Datasheet*.

[6] **Vishay Semiconductors.** *4N35 Optocoupler, Phototransistor Output Datasheet*.

[7] **Infineon Technologies.** *IRLZ34 N-Channel Power MOSFET Datasheet*.

[8] Wescott, Tim. "PID Without a PhD," Embedded Systems Programming, 2000. (Primary resource for integer-math PID implementation).

[9] Perera, Dr. D.G. ECE 5330 Embedded Systems Design Lecture Notes, Fall 2024.

[10] Course Lab Manuals. ESD Lab 6-7: PID Motor Controller, Fall 2024.

[11] D.G. Perera and K.F. Li, "Analysis of Single-Chip Hardware Support for Mobile and Embedded Applications," in Proc. of IEEE Pacific Rim Int. Conf. on Communication, Computers, and Signal Processing, (PacRim'13), pp. 369-376, Victoria, BC, Canada, August 2013.

[12] D.G. Perera and K.F. Li, "Analysis of Computation Models and Application Characteristics Suitable for Reconfigurable FPGAs", in Proc. of 10th IEEE Int. Conf. on P2P, Parallel, Grid, Cloud, and Internet Computing, (3PGCIC'15), pp. 244-247, Krakow, Poland, Nov. 2015.

[13] S.N. Shahrouzi and D.G. Perera, "Optimized Hardware Accelerators for Data Mining Applications on Embedded Platform: Case Study Principal Component Analysis," Elsevier Journal on Microprocessor and Microsystems (MICPRO), vol. 65, pp. 79-96, March 2019.

[14] D.G. Perera and K.F. Li, "Embedded Hardware Solution for Principal Component Analysis," in Proc. of IEEE Pacific Rim Int. Conf. on Communication, Computers, and Signal Processing, (PacRim'11), pp.730-735, Victoria, BC, Canada, August 2011.

[15] D.G. Perera and Kin F. Li, "Hardware Acceleration for Similarity Computations of Feature Vectors," IEEE Canadian Journal of Electrical and Computer Engineering, (CJECE), vol. 33, no. 1, pp. 21-30, Winter 2008.

[16] D.G. Perera and K.F. Li, "On-Chip Hardware Support for Similarity Measures," in Proc. of IEEE Pacific Rim Int. Conf. on Communication, Computers, and Signal Processing, (PacRim'07), pp. 354-358, Victoria, BC, Canada, August 2007.

[17] K.F. Li and D.G. Perera, "An Investigation of Chip-Level Hardware Support for Web Mining," in Proc. of IEEE Int. Symp. on Data Mining and Information Retrieval, (DMIR'07), pp. 341-348, Niagara Falls, ON, Canada, May 2007.

[18] K.F. Li and D.G. Perera, "A Hardware Collective Intelligent Agent", Transactions on Computational Collective Intelligence, LNCS 7776, Springer, pp. 45-59, 2013.

[19] J.R. Graf and D.G. Perera, "Optimizing Density-Based Ant Colony Stream Clustering Using FPGA-Based Hardware Accelerator", in Proc. Of IEEE Int. Symp. on Circuits and Systems (ISCAS'23), 5-page manuscript, Monterey, California, May 2023.

[20] D.G. Perera, "Chip-Level and Reconfigurable Hardware for Data Mining Applications," PhD Dissertation, Department of Electrical & Computer Engineering, University of Victoria, Victoria, BC, Canada, April 2012.

[21] S. Navid Shahrouzi, "Optimized Embedded and Reconfigurable Hardware Architectures and Techniques for Data Mining Applications on Mobile Devices", PhD Dissertation, Department of Electrical & Computer Engineering, University of Colorado Colorado Springs, December 2018.

[22] J. Graf, "Optimizing Density-Based Ant Colony Stream Clustering Using FPGAs", MSc Thesis, Department of Electrical & Computer Engineering, University of Colorado Colorado Springs, CO, USA, March 2022.

[23] A.K. Madsen and D.G. Perera, "Efficient Embedded Architectures for Model Predictive Controller for Battery Cell Management in Electric Vehicles", EURASIP Journal on Embedded Systems, SpringerOpen, vol. 2018, article no. 2, 36-page manuscript, July 2018.

[24] A.K. Madsen, M.S. Trimboli, and D.G. Perera, "An Optimized FPGA-Based Hardware Accelerator for Physics-Based EKF for Battery Cell Management", in Proc. of IEEE Int,l Symp, on Circuits and Systems, (ISCAS'20), 5-page manuscript, Seville, Spain, May 2020.

[25] A.K. Madsen and D.G. Perera, "Towards Composing Efficient FPGA-Based Hardware Accelerators for Physics-Based Model Predictive Control Smart Sensor for HEV Battery Cell Management", IEEE ACCESS, (Open Access Journal in IEEE), pp. 106141-106171, 25th September 2023.

[26] A.K. Madsen and D.G. Perera, "Composing Optimized Embedded Software Architectures for Physics-Based EKF-MPC Smart Sensor for Li-Ion Battery Cell Management", Sensors, MDPI open access journal, Intelligent Sensors Section, 21-page manuscript, vol. 22, no. 17, 26th August 2022.

[27] A.K. Madsen, "Optimized Embedded Architectures for Model Predictive Control Algorithms for Battery Cell Management Systems in Electric Vehicles"; PhD Dissertation, Department of Electrical & Computer Engineering, University of Colorado Colorado Springs, August 2020.

[28] D. Abillar, "An FPGA-Based Linear Kalmann Filter for a Two-Phase Buck Converter Application", MSc Thesis, Department of Electrical & Computer Engineering, University of Colorado Colorado Springs, CO, USA, April 2024.

[29] A. Alkamil and D.G. Perera, "Efficient FPGA-Based Reconfigurable Accelerators for SIMON Cryptographic Algorithm on Embedded Platforms", in Proceedings of the IEEE International Conferences on Reconfigurable Computing and FPGAs, (ReConFig'19), 8-page manuscript, Cancun, Mexico, December 2019.

[30] A. Alkamil and D.G. Perera, "Towards Dynamic and Partial Reconfigurable Hardware Architectures for Cryptographic Algorithms on Embedded Devices", IEEE Access, Open Access Journal in IEEE, vol. 8, pp: 221720 – 221742, 10th December 2020.

[31] A. Alkamil, "Dynamic Reconfigurable Architectures to Improve Performance and Scalability of Cryptosystems on Embedded Systems", PhD Dissertation, Department of Electrical & Computer Engineering, University of Colorado Colorado Springs, 5th February 2021.

[32] M.A. Mohsin and D.G. Perera, "An FPGA-Based Hardware Accelerator for K-Nearest Neighbor Classification for Machine Learning on Mobile Devices", in Proceedings of the IEEE/ACM International Symposium on Highly Efficient Accelerators and Reconfigurable Technologies, (HEART'18), 6-page manuscript, Toronto, Canada, June 2018.

[33] S. Ramadurgam and D.G. Perera, "An Efficient FPGA-Based Hardware Accelerator for Convex Optimization-Based SVM Classifier for Machine Learning on Embedded Platforms", Electronics, MDPI open access journal, 36-page manuscript, vol. 10, no. 11, 31st May 2021.

[34] S. Ramadurgam and D.G. Perera, "A Systolic Array Architecture for SVM Classifier for Machine Learning on Embedded Devices", in Proc. of IEEE Int. Symp. on Circuits and Systems (ISCAS'23), 5-page manuscript, Monterey, California, May 2023.

[35] Jordi P. Miró, Mokhles A. Mohsin, Arkan Alkamil and Darshika G. Perera, "FPGA-based Hardware Accelerator for Bottleneck Residual Blocks of MobileNetV2 Convolutional Neural Networks", in Proceedings of the IEEE Mid-West Symposium on Circuits and Systems (MWCAS'25), 5-page manuscript, Lansing MI, August 2025.

[36] M. A. Mohsin, "An FPGA-Based Hardware Accelerator for K-Nearest Neighbor Classification for Machine Learning", MSc Thesis, Department of Electrical & Computer Engineering, University of Colorado Colorado Springs, CO, USA, December 2017.

[37] S. Ramadurgam, "Optimized Embedded Architectures and Techniques for Machine Learning Algorithms for On-Chip AI Acceleration", PhD Dissertation, Department of Electrical & Computer Engineering, University of Colorado Colorado Springs, 12th February 2021.

[38] J. P. Miro, " FPGA-Based Accelerators for Convolutional Neural Networks on Embedded Devices", MSc Thesis, Department of Electrical & Computer Engineering, University of Colorado Colorado Springs, CO, USA, May 2020.

[39] J. Nurmi and D.G. Perera, "Intelligent Cognitive Radio Architecture Applying Machine Learning and Reconfigurability" in Proc. of IEEE Nordic Circuits and Systems (NorCAS'21) Conf., 6-page manuscript, Oslo, Norway, October 2021.

[40] Kevin Young and Darsika G. Perera, "High-Level Synthesis Based FPGA Accelerator for GPS Signal Image Feature Extraction", in Proceedings of the IEEE Mid-West Symposium on Circuits and Systems (MWCAS'25), 5-page manuscript, Lansing MI, August 2025.

[41] D.G. Perera, "Reconfigurable Architectures for Data Analytics on Next-Generation Edge-Computing Platforms", Featured Article, IEEE Canadian Review, vol. 33, no. 1, Spring 2021. DOI: 10.1109/MICR.2021.3057144.

[42] M.A. Mohsin, S.N. Shahrouzi, and D.G. Perera, "Composing Efficient Computational Models for Real-Time Processing on Next-Generation Edge-Computing Platforms" IEEE ACCESS, (Open Access Journal in IEEE), 30-page manuscript, 13th February 2024.
[43] Mokhles A. Mohsin, and Darshika G. Perera, "High-Level Synthesis-Based FPGA Hardware Architecture for PCA+SVM for Real-Time Processing on Edge Computing Platforms" IEEE ACCESS, (Open Access Journal in IEEE), 24-page manuscript, 18th December 2025. DOI: 10.1109/ACCESS.2025.364576.
[44] Laura H. Garcia, Arkan Alkamil, Mokhles A. Mohsin, Johannes Menzel and Darshika G. Perera, "FPGA-Based Hardware Architecture for Sequence Alignment by Genetic Algorithm", in Proceedings of the IEEE International Symposium on Circuits and Systems (ISCAS'25), 5-page manuscript, London, UK, May 2025.
[45] L. H. Garcia, "An FPGA-Based Hardware Accelerator for Sequence Alignment by Genetic Algorithm", MSc Thesis, Department of Electrical & Computer Engineering, University of Colorado Colorado Springs, CO, USA, December 2019.
[46] R.K. Chunduri and D.G. Perera, "Neuromorphic Sentiment Analysis Using Spiking Neural Networks", Sensors, MDPI open access journal, Sensing and Imaging Section, 24-page manuscript, vol. 23, no. 7701, 6th September 2023.
[47] S. Sharma and D. G. Perera, "Analysis of Generalized Hebbian Learning Algorithm for Neuromorphic Hardware Using Spinnaker" 8-page manuscript, https://doi.org/10.48550/arXiv.2411.11575
[48] R. Raghavan and D.G. Perera, "A Fast and Scalable FPGA-Based Parallel Processing Architecture for K-Means Clustering for Big Data Analysis", in Proceedings of the IEEE Pacific Rim International Conference on Communications, Computers, and Signal Processing, (PacRim'17), pp. 1-8, Victoria, BC, Canada, August 2017.
[49] D.G. Perera and Kin F. Li, "Parallel Computation of Similarity Measures Using an FPGA-Based Processor Array," in Proceedings of 22nd IEEE International Conference on Advanced Information Networking and Applications, (AINA'08), pp. 955-962, Okinawa, Japan, March 2008.
[50] R. Raghavan, "A Fast and Scalable Hardware Architecture for K-Means Clustering for Big Data Analysis", MSc Thesis, (Supervisor Dr. Darshika G. Perera), Department of Electrical & Computer Engineering, University of Colorado Colorado Springs, CO, USA, May 2016.
[51] D.G. Perera and K.F. Li, "A Design Methodology for Mobile and Embedded Applications on FPGA-Based Dynamic Reconfigurable Hardware", International Journal of Embedded Systems, (IJES), Inderscience publishers, 23-page manuscript, vol. 11, no. 5, Sept. 2019.
[52] D.G. Perera, "Analysis of FPGA-Based Reconfiguration Methods for Mobile and Embedded Applications", in Proceedings of 12th ACM FPGAWorld International Conference, (FPGAWorld'15), pp. 15-20, Stockholm, Sweden, September 2015.
[53] D.G. Perera and K.F. Li, "Discrepancy in Execution Time: Static Vs. Dynamic Reconfigurable Hardware", IEEE Pacific Rim Int. Conf. on Communications, Computers, and Signal Processing, (PacRim'24), 6-page manuscript, Victoria, BC, Canada, August 2024.
[54] D.G. Perera and Kin F. Li, "FPGA-Based Reconfigurable Hardware for Compute Intensive Data Mining Applications", in Proc. of 6th IEEE Int. Conf. on P2P, Parallel, Grid, Cloud, and Internet Computing, (3PGCIC'11), pp. 100-108, Barcelona, Spain, October 2011.
[55] D.G. Perera and Kin F. Li, "Similarity Computation Using Reconfigurable Embedded Hardware," in Proceedings of 8th IEEE International Conference on Dependable, Autonomic, and Secure Computing (DASC'09), pp. 323-329, Chengdu, China, December 2009.
[56] S.N. Shahrouzi and D.G. Perera, "Dynamic Partial Reconfigurable Hardware Architecture for Principal Component Analysis on Mobile and Embedded Devices", EURASIP Journal on Embedded Systems, SpringerOpen, vol. 2017, article no. 25, 18-page manuscript, 21st February 2017.
[57] S.N Shahrouzi and D.G. Perera, "HDL Code Optimization: Impact on Hardware Implementations and CAD Tools", in Proc. of IEEE Pacific Rim Int. Conf. on Communications, Computers, and Signal Processing, (PacRim'19), 9-page manuscript, Victoria, BC, Canada, August 2019.
[58] I.D. Atwell and D.G. Perera, "HDL Code Variation: Impact on FPGA Performance Metrics and CAD Tools", IEEE Pacific Rim Int. Conf. on Communications, Computers, and Signal Processing, (PacRim'24), 6-page manuscript, Victoria, BC, Canada, August 2024.

[59] S.N. Shahrouzi, A. Alkamil, and D.G. Perera, "Towards Composing Optimized Bi-Directional Multi-Ported Memories for Next-Generation FPGAs", IEEE Access, Open Access Journal in IEEE, vol. 8, no. 1, pp. 91531-91545, 14th May 2020.

[60] S.N. Shahrouzi and D.G. Perera, "An Efficient Embedded Multi-Ported Memory Architecture for Next-Generation FPGAs", in Proceedings of 28th Annual IEEE International Conferences on Application-Specific Systems, Architectures, and Processors, (ASAP'17), pp. 83-90, Seattle, WA, USA, July 2017.

[61] S.N. Shahrouzi and D.G. Perera, "An Efficient FPGA-Based Memory Architecture for Compute-Intensive Applications on Embedded Devices", in Proceedings of the IEEE Pacific Rim International Conference on Communications, Computers, and Signal Processing, (PacRim'17), pp. 1-8, Victoria, BC, Canada, August 2017.

[62] S.N. Shahrouzi and D.G. Perera, "Optimized Counter-Based Multi-Ported Memory Architectures for Next-Generation FPGAs", in Proceedings of the 31st IEEE International Systems-On-Chip Conference, (SOCC'18), pp. 106-111, Arlington, VA, Sep. 2018.